# Trajectory inference for single cell omics

Alexandre Hutton, Jesse G. Meyer

Department of Computational Biomedicine, Cedars Sinai Medical Center, Los Angeles CA 90048, USA

Correspondence to alexandre.hutton@cshs.org or jesse.meyer@cshs.org

**Abstract**

Trajectory inference is used to order single-cell omics data along a path that reflects a continuous transition between cells. This approach is useful for studying processes like cell differentiation, where a stem cell matures into a specialized cell type, or investigating state changes in pathological conditions. In the current article, we provide a general introduction to trajectory inference, explaining the concepts and assumptions underlying the different methods. We then briefly discuss the strengths and weaknesses of different trajectory inference methods. We also describe best practices for using trajectory inference, such as how to validate the results and how to interpret them in the context of biological knowledge. Finally, the article highlights some applications of trajectory inference in single-cell omics research. These applications include studying cell differentiation, development, and disease.

# Introduction

Omic time courses are of particular interest for understanding the molecular changes that lead to the cellular basis for the progression of disease or identifying predictive markers for pathologies. However, most approaches to single-cell research, such as proteomics or transcriptomics, are inherently destructive to the cells of interest, making it impossible to track a cell's changing omic profiles across time. To address this, computational methods have been developed to stitch together separate samples and order them along a trajectory derived from the cell similarities. The ordering, referred to as "pseudotime," simulates the progression of a cell away from a reference cell state, which can have multiple branching paths.

A simple application is cell differentiation: starting from a stem cell differentiating into multiple cell types, the omic profile will change as progenitor cells transition to cells with specialized functions. One pathway for hematopoiesis, for example, differentiates from hemocytoblasts, to myeloid progenitor, and into megakaryocytes or erythrocytes (among others). The omic profiles of hematopoietic stem cells differ from their descendants [1]; one could sample cells along the trajectory and order their omic profiles according to their position in the lineage relative to the hemocytoblasts.

However, it is not usually possible to link omic profiles to specific points in a lineage. Given the heterogeneity and noise of single-cell omic profiles [2], matching an observed profile to a known profile can't be done with certainty. Further, cellular states are not entirely discrete and display continuous transitions. Instead, trajectory inference aims to connect cells by the similarity of their omic states (**Figure 1**). The rationale is that for a given cell type, there will be a distribution of biologically-functional omic profiles, and that as cells transition from one state to another, that distribution of those profiles will shift. One of the core assumptions across trajectory inference methods is that enough cells are sampled so as to capture the whole transition; otherwise, gaps could lead to ambiguous trajectories that could link trajectories between omicly-similar cells that are not part of the same lineage.

There exists reviews on the transcriptomic methods[2], [3] that cover various methods for inferring trajectories; while this review will offer an overview of these methods, the purpose of this paper is to provide an overview of where these methods have been applied, how they can be best applied to future studies, and discuss some pitfalls that could impact the interpretation of their results.

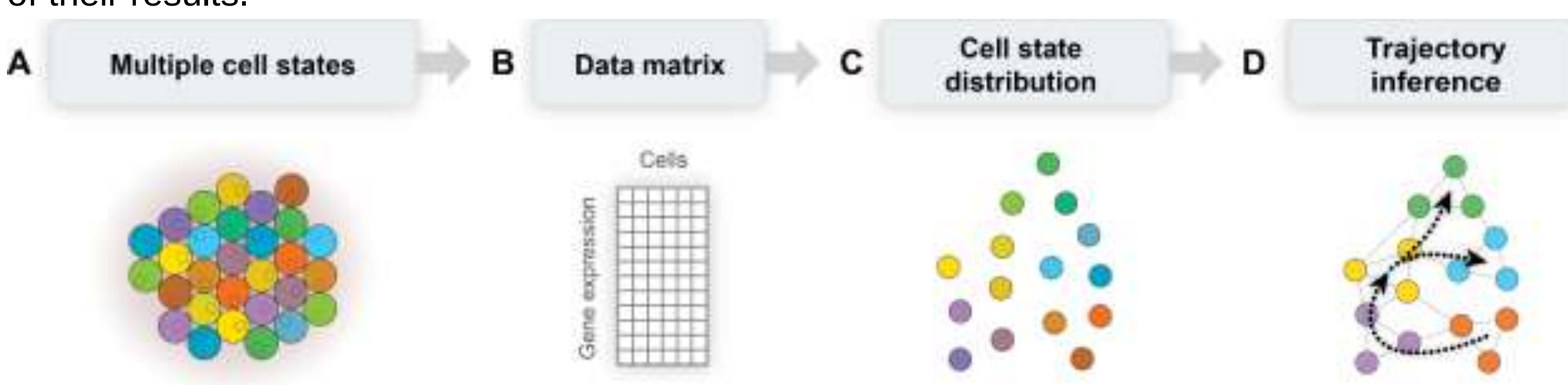

**Figure 1.** Trajectory inference concept. **(A)** A single cell suspension is generated from a complex biological system with multiple cell states. **(B)** Single cell omic data is obtained from the cell suspension, producing a data matrix of quantities related to genes across all measured cells. **(C)** The omic data is used to derive a cell state distribution that is usually portrayed in dimension-reduced space where cells may be clustered by similarity. **(D)** Trajectory inference methods infer the relationships between cells that may indicate development or progression to disease states.

## Continuous State Transitions

The main purpose of trajectory inference is to treat changes in cell characteristics as a continuous process instead of discrete states: the concept of a defined cell state becomes fuzzy at the resolution of single cells. There is not much confusion when discussing differences in a progenitor cell and its descendants, but closely-examining any transition makes it impossible to pinpoint the moment when one discrete state becomes another. Understanding the functional behavior of cells in these transition periods is useful for studying effects such as pathologies progressing differently between patients (e.g. chemoresistance in leukemia [4]) or identifying regulatory genes [5]. Between the unknowable distribution of cell states in single-cell samples and the incomplete information provided by modeling biological processes as discrete transitions, there is a clear advantage to using the continuous transitions obtained by trajectory inference to model changes in cellular function.

# Methods Overview

## Trajectory Inference

There are a variety of tools for performing trajectory inference; Monocle, Slingshot, and PAGA are the most-cited packages (**Figure 2**). We note that Monocle and Slingshot, the two most-cited packages, are made for R, while PAGA and Palantir are implemented in Python. Knowing that R (and more specifically Bioconductor [6]) is widely used in bioinformatics, it would make sense for the packages in R to be adopted more quickly. We also note that Monocle 1-2 is likely to be overrepresented since the Monocle packages require citing their previous versions.
In this section, we offer a summary of how the most-cited methods work in the hope that a better understanding will aid users in knowing what information can be extracted from results and when to be wary.

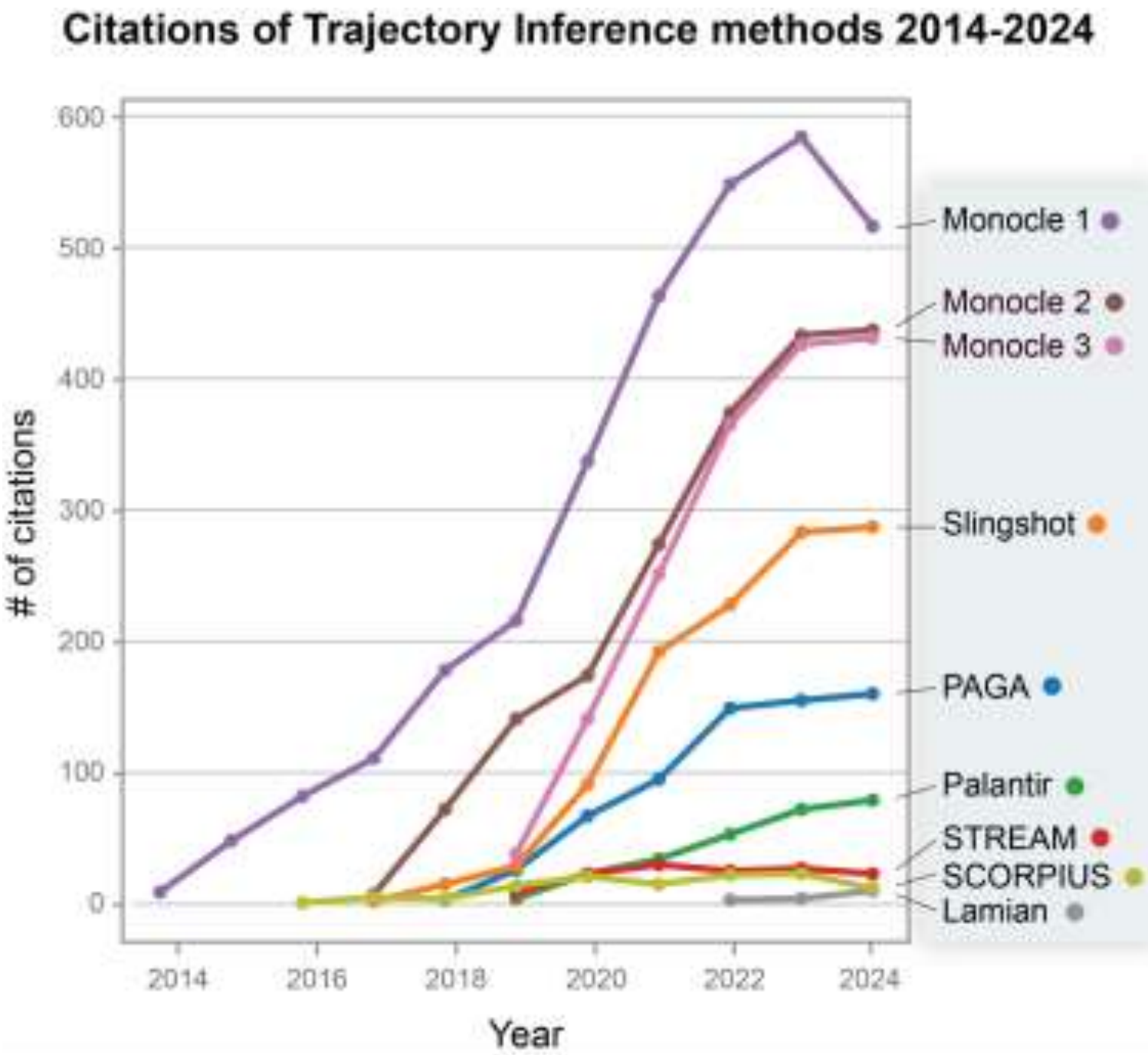

**Figure 2.** Year-to-year citations for the various trajectory inference software packages.

## Monocle (2014, 2017, 2019)

Monocle is a toolkit created for analyzing single-cell RNA-seq data (scRNA-seq), including clustering, differential expression, and trajectory inference. It is currently on its third iteration. Monocle 1 [7] introduced trajectory inference. Monocle 2 [8] improved scalability, particularly by supporting sparse matrices, and updated the trajectory inference to use reversed graph embedding to reconstruct trajectories. The latest iteration, Monocle 3 [9], expanded its applicability to large datasets (on the order of millions of samples) while removing some assumptions made by the underlying implementation and allowing trajectories to better fit the variety of cell behaviours and observed datasets (multiple origins, cell state cycles, converging cell states from different origins). As stated in the Monocle 3 paper, Monocle shifted its approach to using a mix of PAGA (see later section) [10] to scale to larger datasets and using the PAGA clusters to inform finer-resolution trajectories. Monocle projects the high-dimensional scRNA-seq data to a low dimensional state using UMAP[11], which allows for local distances to be computed more efficiently and preserves the global structure of the data better than t-SNE [12]. Clustering is then performed using the Louvain algorithm to identify groups with similar expression patterns. Using these groups, a graph is constructed using a variant of the SimplePPT algorithm [13], [14] to allow it to use principal curves on large datasets and to produce a graph containing loops. From there, pseudotime values are computed by projecting each sample onto the trajectory and computing the distance from a root node. In the case of multiple root nodes, pseudotimes are taken as the minimum distance across all root nodes. See the supplementary GitHub repository for a notebook tutorial of Monocle usage with notes (https://github.com/xomicsdatascience/trajectory_review/).

## Slingshot (2018)

Slingshot [15], introduced by Street *et al.*, uses a method for trajectory inference that is similar to Monocle 1 [7] and 2 [8] , with some key differences that optimize robustness against sources of noise, modularity of use with different pre-processing steps, and better identification of multiple lineage branchings. Monocle uses a minimum spanning tree (MST), where its implementation fits all available data points. When the computed pseudotime for a given dataset was compared to that of a subsampled set, Street *et al.* found that the pseudotime ordering changed dramatically compared to other methods and concluded that the approach was not sufficiently reliable. By contrast, methods using principal curves[16] to compute pseudotime were most stable against subsampling, but are not able to account for branching lineages. To address this, Slingshot computes a MST from clusters in the data instead of using the data directly, and then computes a curve for each trajectory; the MST allows for flexibility of branching lineages while the principal curve offers robustness to noise and generalizability to similar datasets. Once the curves are obtained, pseudotime for each sample is computed by projecting it to the closest curve and evaluating its distance along the curve from the user-specified starting cluster. A marked difference between Slingshot and other approaches is its ability to fit into different workflows; Slingshot assumes that the data is clustered, but it does not require a specific clustering to function. Street *et al.* evaluated different clustering methods with different total clusters and found that although performance decreased quickly at low cluster counts and dropped off gradually as the counts increased, it was comparable across both clustering methods and parameters.

## Partition-based graph abstraction (2019)

Partition-Based Graph Abstraction (PAGA) [10] is an algorithm for discovering relationships present within single-cell RNA-seq (scRNA-seq) data. There are generally two approaches to representing single-cell data: clustering and continuous transitions. While clustering approaches present us with an intuitive interpretation of distinct cell groupings (e.g., cell types) that can be determined by a cell's transcriptome, it has some limitations. Single-cell experiments regularly demonstrate that omic profiles are heterogeneous, even among specific and known cell types; if a cell differentiates into a different cell type, its transcriptome will change continuously to be more like its new type, but it is unclear where along that transition it should be labeled as its new type. It is also unclear how cell similarity can be evaluated reliably. For example, take a cell that differentiates multiple times with a terminal state similar to its starting state (**Figure 3**), evaluating similarity using inter-cluster distances would place the final cell type "closer" to its starting type than its intermediate states. By contrast, continuous approaches to identifying connections paint a richer picture of the variability of omic profiles, but make some assumptions about the data distribution that don't generally hold as follows. The core assumption is that the underlying processes are accurately captured by the data, which is less likely to be true for rapid transitions. In cases where the process is not sufficiently well-sampled, the trajectory between cell states could be misrepresented. Continuous trajectory inference methods are also ill-equipped to deal with unconnected sections of the data. If multiple unrelated cell types are present in the data, these methods, the assumption is that these should nonetheless be connected on the same trajectory. This is particularly problematic when dealing with in vivo samples where multiple cell types are likely to be present despite having separate lineage and functionality.

PAGA combines the clustering and continuous approach. Although previous methods also represent the data as graphs, single-cell data's noisy nature and sparse sampling make it difficult to determine which cells should be grouped together. By using a multi-resolution approach to creating the graph and statistical model for node (cluster) connectivity, PAGA is able to accommodate data distributions that are more in line with single-cell data, including disconnected clusters, sparse sampling, and continuous changes between cell states. See the supplementary GitHub repository with a tutorial notebook demonstrating and explaining the usage of PAGA (https://github.com/xomicsdatascience/trajectory_review/).

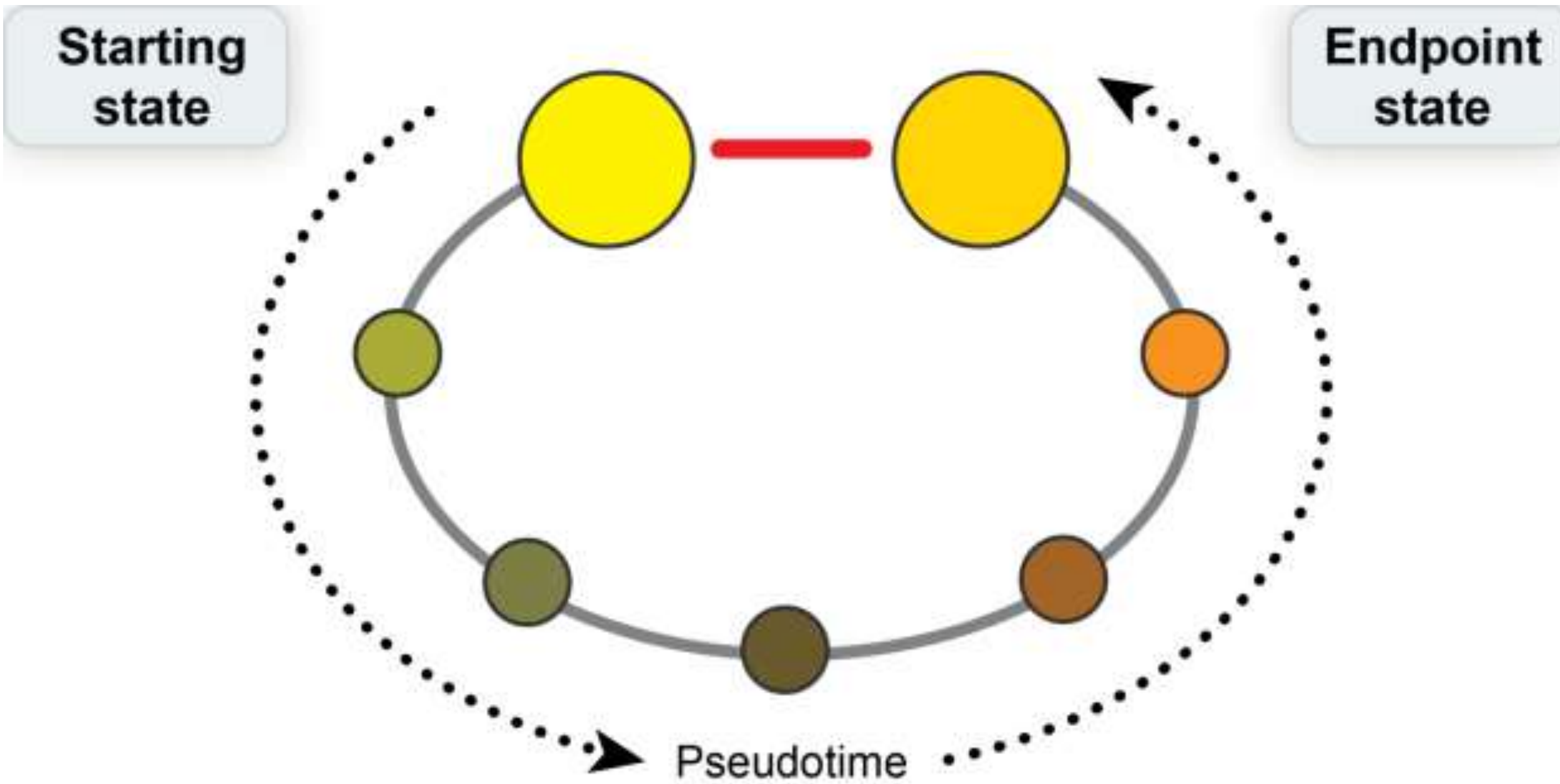

**Figure 3.** Schematic of inter-cluster distances in 2D space. By judging the space using only the distances in gene space between clusters, we might conclude that the two top-most clusters are linked. Instead, treating the process as continuous and inferring a trajectory would reveal that there are intermediate stages through which cells transition before reaching the terminal state.

## Palantir (2019)

Palantir [17] is an approach that treats cell trajectories as primarily continuous, whereas previous work had treated them as transitions between discrete states. Since single-cell omic studies have shown that cell transitions appear to steadily shift from the omic profile of one state towards the profile of the other, and that since discrete borders are not immediately apparent, it is reasonable to treat cell types as a continuum.

To compute trajectories, Palantir first places the data into a low-dimensional space using a kNN graph using diffusion maps (**Figure 4**) [18]. Diffusion maps can be made to better model the similarity between cells by including a similarity function that transforms the distance in some way; based on previous work, Palantir uses an adaptive Gaussian kernel to transform the Euclidean distance. Typically, a Gaussian kernel has a fixed width (variance) and penalizes the similarity of two cells exponentially as a function of the Euclidean distance between them. Knowing that the density of cells along a trajectory is usually variable (more cells in stable states, fewer in transition areas), Palantir opts to use an adaptive kernel whose width varies with the density of samples, allowing for variable densities and making trajectories more stable. To compute pseudotime along these trajectories, the direct approach of accumulating shortest-path distances from the origin to other cells has been found to accumulate noise as the total distance increases. To limit this impact, distances are instead computed from the nearest waypoint, which are distributed uniformly along the estimated trajectories. The core to Palantir's approach is to then use these waypoints as states in a **Markov chain**, establish directionality using the computed pseudotime, and identify terminal states. **Markov chains** model the transition from one state to another using a transition probability matrix, and that transition matrix depends only on the current state; this states that all information from previous states that might affect the transition matrix is also present in the current state. The Markov chain is directly comparable to the underlying cells. Finally, the group defines the **differentiation potential** for each state within the Markov chain as the entropy of each state computed using the probabilities of a state reaching different terminal states. In the context of modeling cell differentiation, entropy is *high* when there are multiple terminal states that can be reached from the current state, and entropy is *low* when there are fewer possibilities. Differentiation potential reflects how well one can predict where the cell will end up; high potential is low predictability and vice versa.

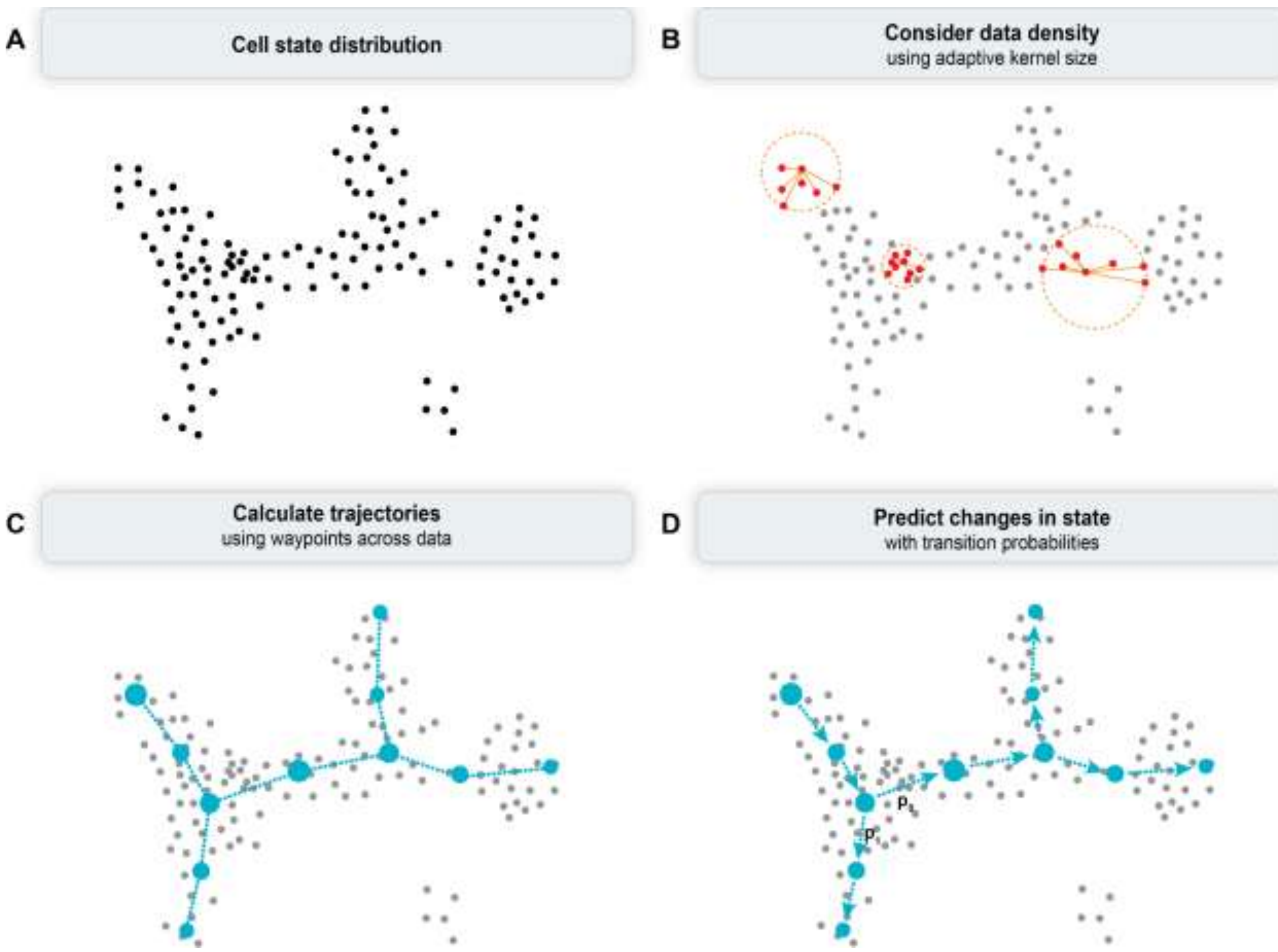

**Figure 4.** Steps used by Palantir for determining trajectories. **(A)** Distribution of data sample. **(B)** Using an adaptive Gaussian kernel, the distance between cells is obtained and used to obtain a diffusion map. **(C)** Using the obtained distances, waypoints are distributed evenly throughout the space. **(D)** Transition probabilities are computed, modeling the chance that a cell in one state transitions to downstream states. These probabilities are later used to compute the entropy of a state.

## Interactive tutorial notebooks

Tutorials for Monocle using R and PAGA using Python are available on GitHub (https://github.com/xomicsdatascience/trajectory_review/). Using the data for the Paul *et al.* 2015 [19] paper commonly used in Scanpy [20], we demonstrate how to process data and obtain trajectories (**Figure 5**). Note that although there is a general agreement between trajectories, the pseudotime estimates[1] do differ. We recommend that users that have not used trajectory inference before to start by going through these notebooks to gain an understanding of each step, and then apply the methods to their own data.

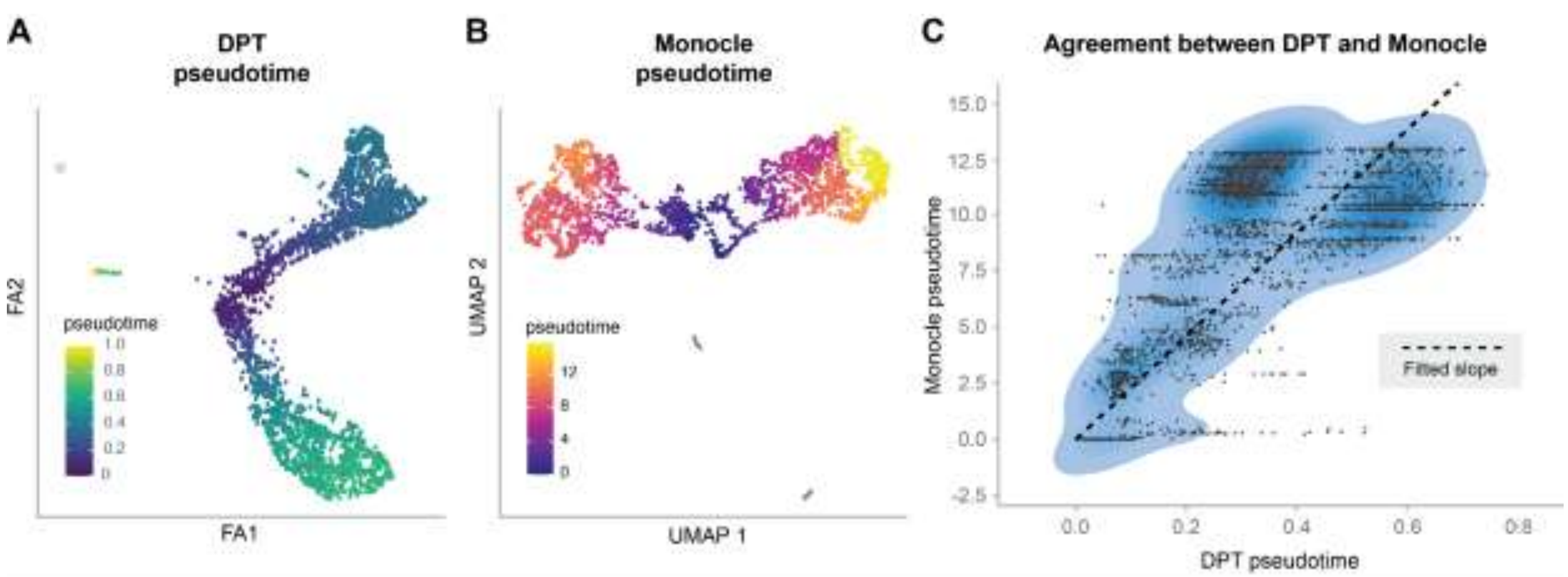

**Figure 5.** Pseudotime estimates computed for the same dataset using different methods. **(A)** Diffusion pseudotime computed with Scanpy. Data is projected on Force Atlas axes, a method for determining a graph layout. **(B)** Pseudotime computed by Monocle 3 projected onto UMAP axes.

[1] For PAGA, pseudotime estimates were obtained using diffusion pseudotime [21], [10].

# Computing Pseudotime

Although related, trajectory inference and pseudotime are different methods. Whereas trajectory inference is used to estimate how cell states change across gene space, pseudotime is a distance metric that measures distances relative to some origin. Pseudotime should be considered as a measure of omic change along some trajectory rather than a measure of time, and external information is required to connect pseudotime to actual time or some kind of progression. At its simplest, one could compute the Euclidean distance between an origin and all other samples and this would constitute a valid value for pseudotime by definition, though not likely a useful value. Inferred trajectories are useful for making more reliable estimates of pseudotime; distances for a given sample represent the distance between the origin and a point on the trajectory to which a sample is mapped. Some of the methods discussed in the previous section implicitly compute pseudotime when returning a trajectory. Monocle [9] first computes a trajectory for a given dataset and projects each sample onto its nearest point on the trajectory. From there, a sample's pseudotime is the geodesic distance along that trajectory; that is, the sum of the distances between successive points that are on that trajectory starting from a specified root. By contrast, methods such as diffusion pseudotime (DPT) [21] compute pseudotime by selecting a root, applying a distance kernel to estimate distances and obtaining transition probabilities (the probability of a sample transitioning to a nearby state).

## Diffusion Pseudotime

Diffusion pseudotime (DPT) [21] models changes in cell state as diffusions. Weighted distances between samples are obtained using a Gaussian kernel, and the probability of a cell transitioning from its current state to another is obtained from random walks across all cells. By considering multiple paths, DPT is able to obtain a distance metric that is more robust to changes in preprocessing or subsampling. Although it is possible to obtain branching trajectories by post-processing the results, the authors recommend against using it in favor of PAGA [22].

## psupertime (2022)

One of the major limitations for the previously-described methods is that they are unsupervised; pseudotime ordering is determined by distance metrics computed in a gene expression subspace. This approach can be useful when examining a biological process where no other information is available, but it can be misleading in cases where multiple biological processes are affecting an omic layer. When pseudotime reflects more than one process, assigning the behavior of genes to the process of interest is no longer possible. However, it is not uncommon for experiments to contain metadata or information about an experiment that can help inform cell ordering. For instance, drug response experiments have set time points where samples are collected; incorporating the time step into the ordering can ignore sources of noise that would confuse the order of cells. Methods such as psupertime from Macnair et al. [23] include this information into the ordering by treating the data as labels. In the drug response example, time steps 1-5 would be treated as separate ordinal categories, and logistic regression models are fitted to the data to predict the labels. The model is L1-regularized to identify a small subset of genes that are predictive of the ordering, then uses those genes to place cell samples in pseudotime. Although the method can be used with any experiment that has ordered labels, it also assumes that those ordered labels are related to a single process. Processes that consist of multiple branches (e.g., different drug reactions) would have the same time points, and psupertime would not differentiate between them and, as shown in their synthetic experiments, the selected genes would reflect changes common to all branches and not branch-specific ones. To make the best use of psupertime, its authors recommend first separating possible branches through the use of trajectory inference, then applying psupertime to identify predictive genes.

### Chronocell (2025)

Fang et al. [24] suggest a more principled approach for determining cell ordering. The method discusses the limitations of trajectory inference and how pseudotime methods lack a ground truth to compare their results. Chronocell models trajectories using state transitions, including a latent variable to represent "process time" that corresponds to the progress of a cell along the biological process. The obtained pseudotime is explicitly tied to the model of the trajectory, but that is implicitly the case with other approaches and should not be seen as a limitation. The approach is most useful in instances where the topology for a process is sufficiently well-known that it can be directly specified a priori.

## General Assumptions

Although each method makes separate assumptions in their approaches, there are general assumptions that are common across the different methods.

**Sampling**. The biological process being observed should be sufficiently sampled so that all parts are represented in the collected dataset, including transition states (**Figure 6)**. Given the variety of processes, it is not possible to prescribe specific numbers. Tools such as SCOPIT [25] can offer estimates for the required number of samples when seeking rare cell states, but since one does not usually have values for the necessary inputs, the values they produce may not reflect the best experimental protocol. A discussion of statistical power considerations in single-cell experiments has been previously published by Jeon et al. [26]. The limitation is present even when observing known processes; cell transition states (e.g., partway through differentiation) will probably be less common than their terminal states. Knowing whether a sufficient number of samples has been collected is likely to be iterative in practice and part of developing an experimental protocol; the important criterion is that there are no large gaps present in the trajectory (and by extension, in pseudotime). To our knowledge, no analysis of what presents a reasonable and acceptable gap has been done, and inference methods will usually use some heuristic (Palantir uses the distance of the $n^{th}$ nearest neighbor, PAGA uses interconnectedness) that has been found to work across different datasets. There is an inherent weakness in this approach: iteratively acquiring data in pursuit of a desirable outcome is tantamount to p-hacking. To avoid this, sample count requirements should be estimated in the first phase of an experiment before the collection of the full dataset in a later phase.

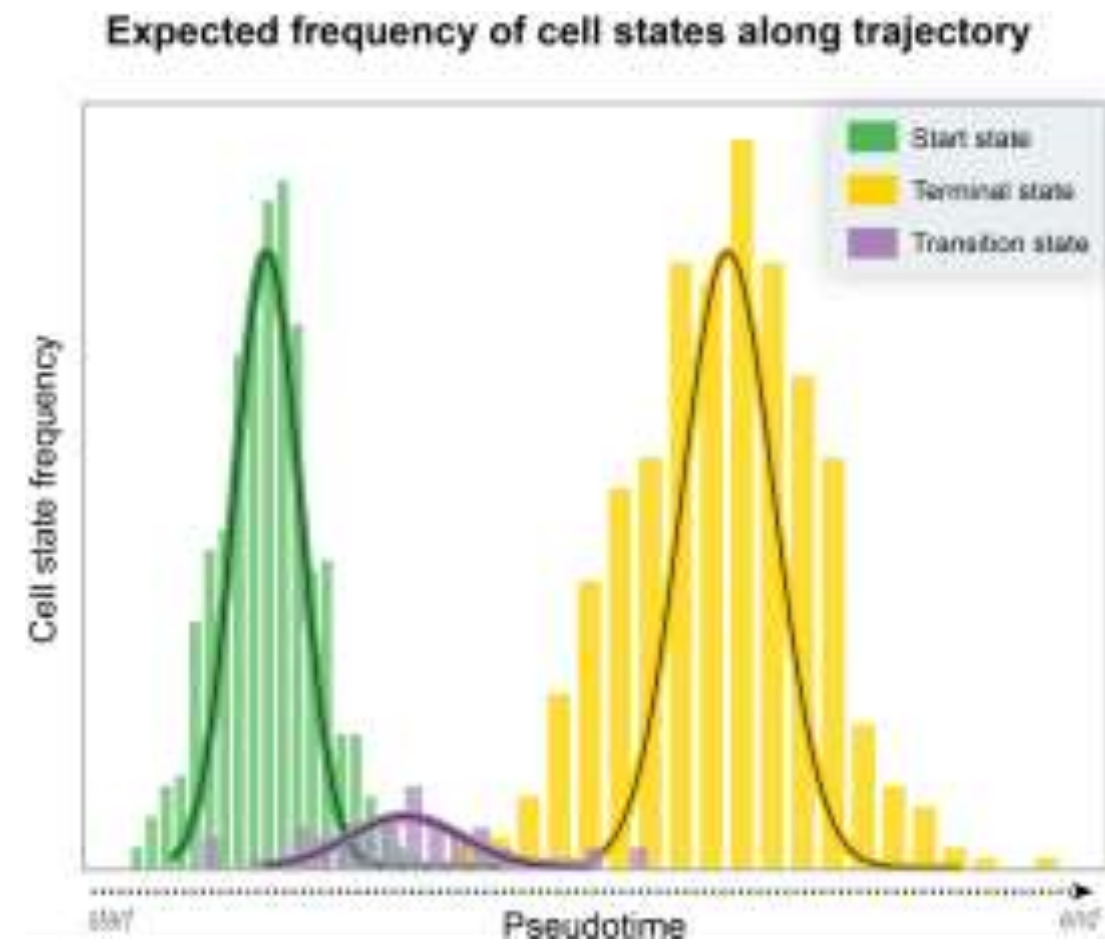

**Figure 6.** When sampling a biological process, experiments must acquire a sufficient number of datapoints to detect rare states; for some processes, transitions may be quick and have a low chance of being

**Continuity**. Although related to sampling, the assumption of continuity does not entirely overlap with sampling. All trajectory inference methods assume that a biological process is continuous in gene-space; their degree of gene expression changes continuously across the process being observed. If this assumption holds, then a sufficiently-sampled process should have cells representing the entire spectrum of cell states. Consequently, we would conclude that cell states that don't have a line of cells between them are not directly related. This assumption is important to consider when examining gene expression across pseudotime; downstream trajectories are user-identified from computed nodes or clusters and are identified due to some biological question being explored (e.g., one could have a hypothesis that cell A differentiates into B then C, and select nodes or clusters that relate to those cells). If the pseudotime across the selected clusters presents a gap, this indicates that either (1) the process is insufficiently sampled, (2) the trajectory is missing an intermediate state, or (3) the cells are not directly related. Examples for (1) are straightforward: the biological process is present, but more samples are required to observe it. For (2), a user could select hematopoietic stem cells and macrophages from their dataset; the data is sufficiently sampled and they are part of the same lineage, but intermediate stages were not selected. Lastly, (3) the selected clusters are not directly connected to one another due to being from entirely different lineages or on different branches from a common ancestor.

**Pseudotime is relative**. At its core, pseudotime is a measure of distance from a starting position. When used to investigate a biological process, it can be treated as a unit-less measure of progress along that process. Some methods normalize it to the [0,1] range to make the extreme values consistent across applications, while other implementations keep the index of the sample order. Consequently, different portions of pseudotime are not directly comparable across datasets; for example, if one dataset contains descendants that another does not, the cell state distributions across pseudotime would shift despite there being no difference in the overlapping cells (**Figure 7**).

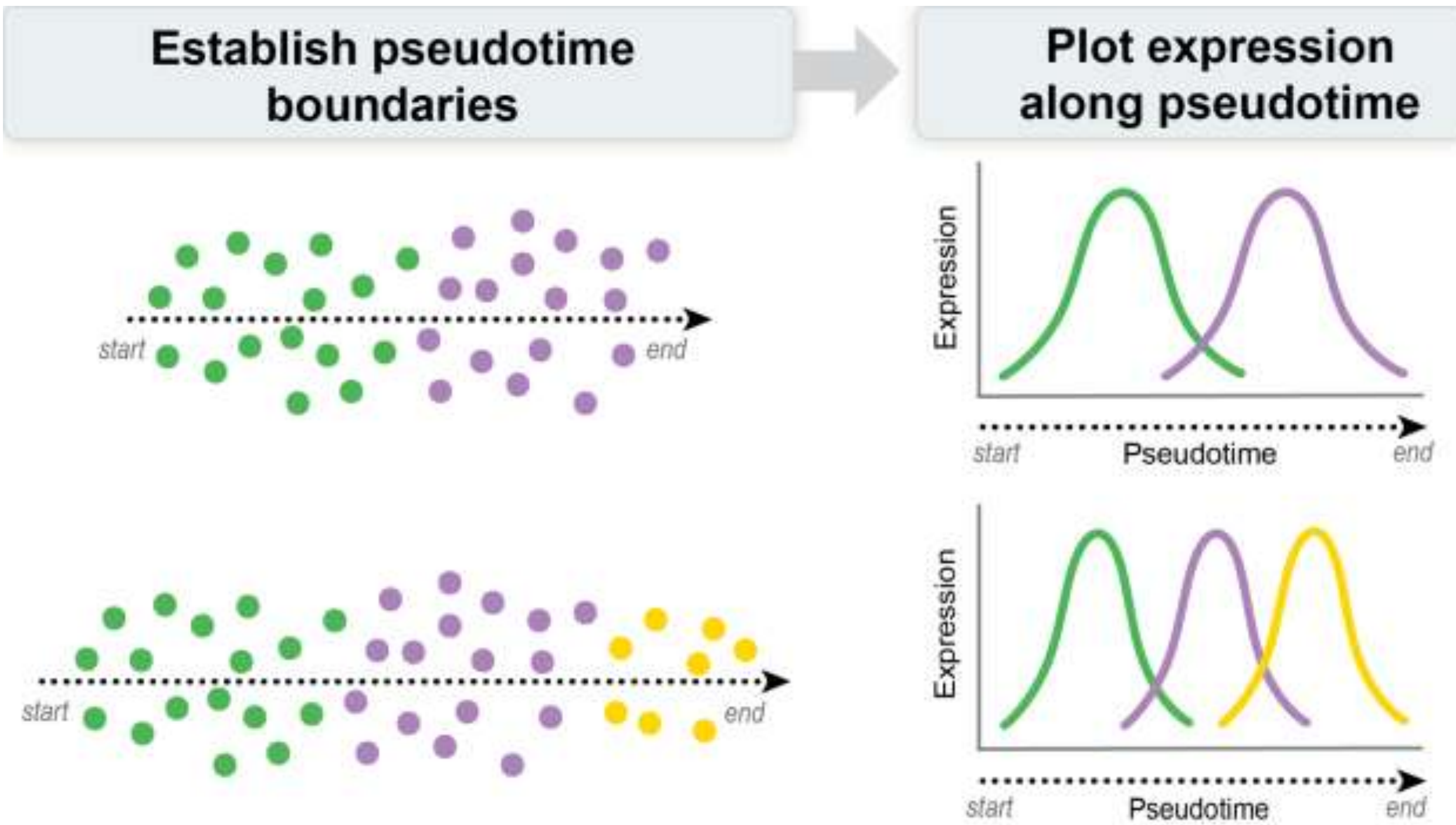

**Figure. 7.** Two almost-identical datasets, with the lower one containing extra cells near the end of the trajectory. Since the trajectories have different endpoints, directly comparing the “middle” or “end” of pseudotime is not meaningful. Methods that don’t normalize pseudotime have a similar problem; adding more samples in one cluster would compress all other clusters in pseudotime.

## Connecting pseudotime to time or progress

Trajectory inference identifies cells that are close to one another in gene space and proposes a path through the dataset to connect different cell states. However, it is not possible for these methods to infer any kind of directionality. Several pieces of external information are required to determine and validate connecting pseudotime to time.

**Origin or root cell**. A starting point is required to identify the directionality; from a root node, one can set the pseudotime to be 0 and increase it along any connected trajectories. Root nodes are typically selected using gene markers, either for stem cells or some otherwise known starting cell state, such as the healthy state. Most methods support only one root, but some methods can accommodate multiple roots, such as Monocle 3 [9] (by taking the minimum across all roots) or DPT [21] (by taking the average).

**Trajectory validation**. When investigating a new biological process, it is important to validate computational findings with known markers to ensure that the trajectories can be used to extract new information. The validation can be done by identifying cell types along a lineage and verifying that they have the expected change in biomarkers (e.g., stem cells expressing marker

A, then shifting into expressing marker B as they differentiate); one would expect a continuous change in expression in those markers as the cells move through the trajectory. Once parts of the trajectories are validated, examining differences between groups (e.g. disease vs. control) and differences between trajectories can be used to identify points of interest. Ultimately, differences would need to be experimentally validated, but ensuring that the obtained trajectories are consistent with known biological function is a simple verification to avoid incorrect trajectories found due to noise or computational artifacts. Subsampling is a useful tool in this regard: by taking subsets of the data, computing a trajectory, and projecting the remaining data onto the trajectory, we can establish the stability of trajectories and evaluate whether we find gene expression trends that are consistent in different subsamples.

**Sampling across time**. Certain biological processes are better understood by population shifts in cell states. Typical fetal development would contain more stem cells and progenitors than cells in a terminal state, and we would expect fewer progenitors as development progressed. Regardless of other markers, the relative frequency of stem cells to terminal states across developmental stages would itself be a marker for development. Determining these shifts is done by first collecting multiple samples across time, each with their own distribution of cell states along development time. Then one can perform trajectory inference and split the samples by their timestamp. Meng *et al.* [27] used this concept by collecting data from multiple timepoints and found changes in cell subpopulations during neonatal swine development, where progenitors increased while goblet and tuft cells decreased. The change in distribution would otherwise not be possible to demonstrate based on a single time point.

## Interpreting Trajectories

Results from trajectory inference lends itself to misinterpretation partly because it is a visually-simple projection of high-dimensional data. We outline some aspects around trajectory inference that are helpful to be mindful of.

**Pseudotime is not time.** Even if a trajectory is connected to time (via time series), it is not directly reflective of it. Pseudotime is a distance in omic space. Suppose that a cell transitions across states A→ B → C and that each state occupies 1/3 of pseudotime. State B is omicly halfway between A and B, but the cell could spend only a small fraction of its time in state B and its position in pseudotime would not inform us of when the transition happens in time. This is particularly important for processes such as drug responses, where relating a trajectory to time could be important for identifying when negative effects become irreversible.

**Projections lose information.** Visualization requires casting complex data into at most three dimensions. Excluding degenerate cases, dimensional reduction necessarily removes information. Three points forming an equilateral triangle (2D) cannot be put on one line (1D) without losing information about the distance between the points. Distances between parts of a trajectory are not immediately meaningful; conversion to pseudotime that considers the full omic distance is important for determining progression along a process. Consequently, visual inspection of trajectories is not sufficiently reliable for determining their plausibility.

**Trajectories may not be stable.** As discussed by Durmaz et al. [28], preprocessing steps have a significant impact on downstream analyses. The choice of normalization, dimension reduction, and pseudotime method have notable effects on the repeatability of results. Analyses that used non-linear dimension reductions such as UMAP [11] or t-SNE [12] are more sensitive to changes in preprocessing [28]. We should then ask: can we draw useful conclusions from results that require a specific set of processing parameters? Perhaps, but to do so would require a way to evaluate the trajectories, and trajectories that are obtained through computational methods typically lack a ground truth to evaluate them. We can mitigate this by running the trajectory inference using multiple sets of parameters, data subsets, and methods, and examining the consistency of the results. When doing so, it is important to ask questions such as, "How reliably are branching lineages identified in this dataset?" or "Do we obtain different pseudotime ordering when using a subset of the data?" If the results are not robust to these changes, then it is important to investigate as to why that is. To that end, we recommend exploring dynverse [29], which provides an interface for multiple R-based trajectory interface methods.

**Trajectories are computational.** Some of the previous points can be consequences of trajectory inference being a computational method: trajectories can be obtained even for meaningless, simulated data. Validating trajectories quantitatively can not be done using the same data that was used to generate the trajectories, as this is at best circular (see Supplementary text of Fang et al. [24]). Further, comparing against known markers occurring at known states would validate those markers, but not other genes. Thus, information obtained from trajectory inference is either redundant because it is consistent with existing knowledge, or requires validation. Ensuring that trajectories are consistent with existing knowledge can rule out trajectories that aren't likely to be useful, which reduces the amount of follow-up work required. The utility of trajectory inference is in identifying previously-unknown genes that are related to a process that would warrant further investigation (e.g. using instrumentation to investigate the specific gene).

# Techniques Enabled With Inferred Trajectories

The computation of cell trajectory relationships opens the ability to perform additional computational analyses. For example, other statistical and information theory techniques like Granger causality and transfer entropy can enable understanding of relationships between molecules across inferred trajectories. Often these approaches attempt to infer gene regulatory networks (GRNs). This area of research is so active that there are robust benchmarking frameworks available [30]. **Figure 8** gives an overview of such applications.

Given a computed pseudotime, we may want to know which genes statistically change along that pseudotime to provide a view of the molecular mechanisms that mediate progression along that pseudotime. Early methods for DE calculation were limited in that they were tethered to specific methods for pseudotime computation, they used generalized additive models and lacked proper estimation of calibrated p-values, or they used general linear models that required invalid assumptions. To solve all these problems, in 2021 PseudotimeDE was introduced, which uses a negative binomial GAM or zero inflated NB-GAM to model uncertainty and improve DE estimation [31]. The paper uses five real-data examples to demonstrate the PseudotimeDE approach. In each of these examples, the authors apply PseudotimeDE along with other DE methods (tradeSeq and Monocle3-DE) and compare their performance based on p-value calibration, power, and functional analysis of the identified genes.

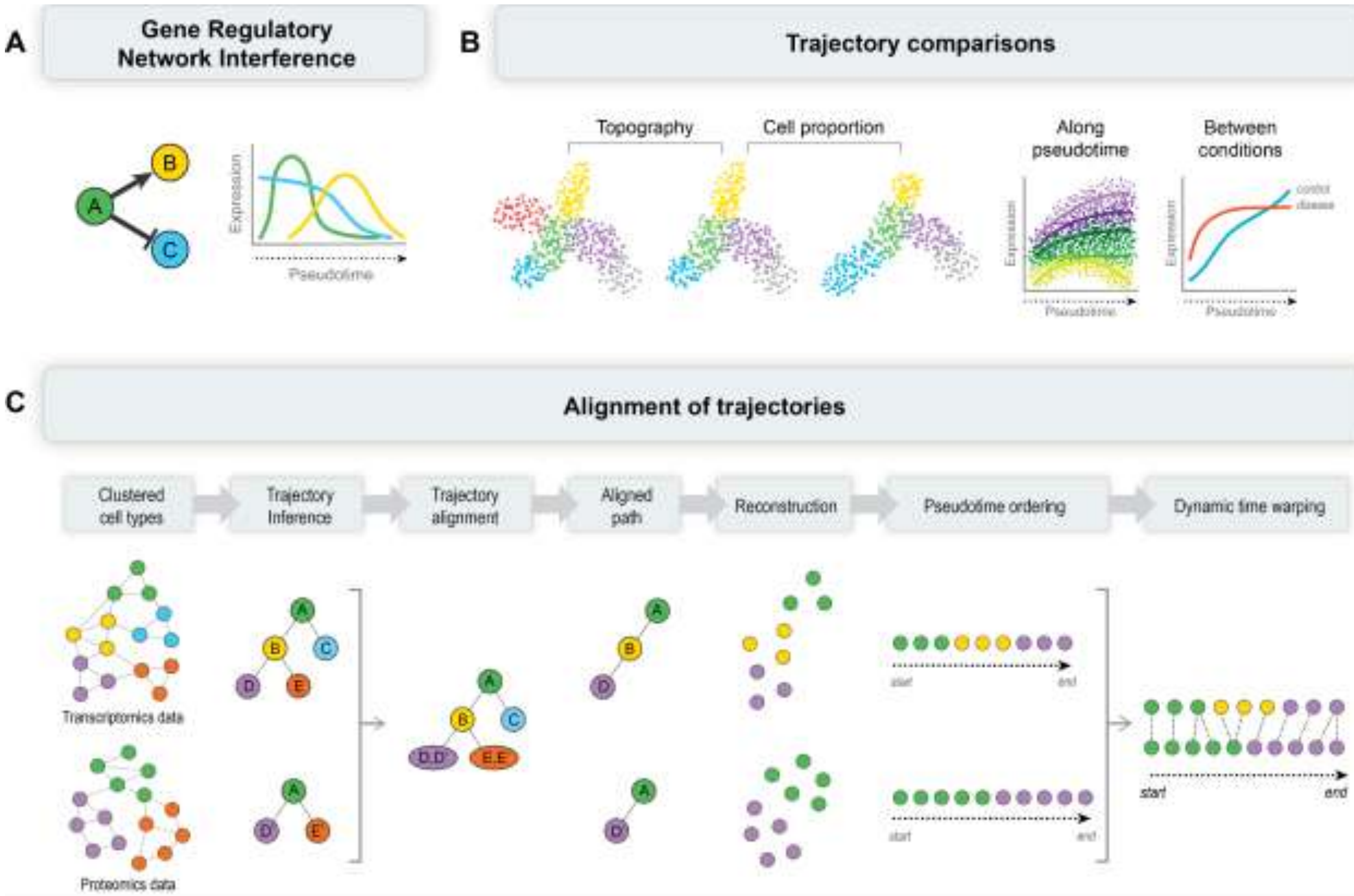

**Figure 8.** Overview of applications of trajectories. **(A)** Inferring gene regulation networks from the distribution of gene expressions across pseudotime. **(B)** Comparison of inferred trajectories between two conditions, including the graph layout of the different conditions or population shifts. **(C)** Different omic measurements can result in different trajectories; the graphs can be aligned and combined to give a better understanding of the process being analyzed.

Granger causality estimates whether one time series is causative of another time series. Combining this concept with inferred trajectories of single cell omics data, Granger causality could estimate which molecules influence other molecules over pseudotime. However, due to issues mentioned above related to sparsity in trajectory space and also in missing values of measured molecules, the application of Granger causality to single cell omics trajectories is not trivial. Deshpande *et al.* introduced an approach called SINGE, which is based on kernel-based Granger causality regression that deals with these issues [5]. Importantly, this approach averages an ensemble of regressors to construct a ranked list of regulator-gene interactions. The authors showed superior area under the precision recall curve for detecting known gene regulatory networks from single cell data relative to competing approaches using ESC-to-endoderm differentiation data. The authors further demonstrated good performance of GRN discovery from retinoic acid driven differentiation of mouse embryonic stem cells.

Transfer entropy is a technique from information theory that measures how much information is transferred between two time series datasets. Like Granger causality, transfer entropy also can enable inference of relationships, and is thus another natural analysis technique that can be used to understand the relationships between molecules within an inferred single cell omics trajectory. Kim *et al.* first demonstrated this concept in 2020 with their software TENET [32]. The found superior performance to other GRN inference algorithms when applied to mouse embryonic stem cell differentiation, and also when examining direct reprogramming of mouse fibroblasts into cardiomyocytes. They extensively benchmarked TENET against competitors using Beeline, and found generally very strong performance in predicting targets of transcription factors Nanog, Pou5f1, Esrrb and Tbx3. TENET was even able to discern context specific GRNs, where Nme2 was required for differentiation in 2iL media but not in SL media. Another group recently developed a faster GPU-based implementation of TENET called FastTENET[33].

Given pseudotime trajectories, another task of interest may be to compare trajectories across multiple biological conditions. Hou *et al.* developed the Lamian framework for making statistically rigorous comparisons across groups [34]. The authors claim to overcome two main problems. First their framework can detect a diversity of potential changes in trajectories, which can include (1) topology differences, (2) cell proportion differences, or (3) gene differential expression (DE) changes across mapped cells. Second, their framework can account for separation of biological and technical signals. Their framework includes preprocessing, tree variability estimation, and differential tests along trajectory for gene changes. The authors demonstrated the utility of their package applied to scRNA-seq from bone marrow samples and then applied it to discover differential gene activation in T-cell activation related to COVID-19 infection.

CAPITAL (Comparative Analysis of Pseudotime Trajectory Inference with Tree Alignment) is a valuable new method for comparing single-cell trajectories, even when those trajectories include branches [35]. Trajectory alignment across single cell omics datasets may be valuable for several reasons, for example comparing the dynamics of gene expression across species, comparing trajectories generated separately for across control and perturbation to discover differential regulators, detecting cell type diversity changes between conditions, or for understanding multi-omic trajectory differences in data measured from parallel cell populations. Existing tools primarily focus on aligning single lineages across datasets with dynamic time warping, which makes them unsuitable for comparing branching trajectories. CAPITAL addresses this limitation by first clustering cells in each dataset and then using a dynamic programming algorithm to perform a tree alignment of the clusters across datasets. Once an alignment of the trajectory trees has been obtained, the single cells within each aligned cluster are aligned using dynamic time warping. This approach enables the comparison of branching trajectories in a global fashion without requiring prior knowledge of the linear paths to be compared. The authors demonstrated this approach by aligning trajectories of blood cell differentiation across species and found that common markers of differentiated blood cells, such as IRF7, CSF1, and ELANE, have different magnitudes of change between human and mouse. CAPITAL offers a promising new way to investigate the dynamics of cellular processes such as differentiation, reprogramming, and cell death, as well as to identify regulators of human disease.

# Applications of Trajectory Inference

Trajectory inference is a useful tool that is applicable to a wide range of experimental settings. In this section, we briefly discuss some notable papers,their settings, and how their findings are enhanced by trajectory inference. The later section covers some papers that used trajectory inference but either did not benefit from it, over-interpreted some results, or could be potentially misleading to those unfamiliar with trajectory inference. For those papers, we restrict our discussion to the use of trajectory inference methods and note that it represents only one and often small part of the work; it should not be taken to reflect the overall merit of the findings reported in the papers. The papers discussed here are summarized in Table 1.

## Case studies of notable utility

**Integrating external information.** A consistent pattern among the strongest applications is the integration of non-omic data to verify whether trajectories were consistent with the known progression of a biological process as demonstrated by Xin et al. [36], Kirita et al. [37], Mukherjee et al. [38], Smajić et al. [39], and Kannan et al. [40] Examples include aligning pseudotime with Braak stage or CERAD score in Alzheimer's disease [38], matching pseudotime to injury timepoints in acute kidney injury [37], and confirming root placement and ordering by distance to known pathological or healthy end states [39]. Some studies establish the directionality of trajectories by combining pseudotime with RNA velocity to confirm they are biologically plausible [40]. These strategies reduce overinterpretation and strengthen the argument that the trajectory captures real biological progress.

**Comparing trajectories to known biology.** Another positive pattern appears in studies that begin by confirming the trajectory structure using established biological expectations, as seen in Oprescu et al. [41], Yoshimura et al. [42], Lotto et al. [43], Chen et al. [44], Zhang et al. [45], Zhang et al. [46], Tosti et al. [47], and Li et al. [4]. These analyses reproduce known cell-state transitions, such as immune activation and resolution during muscle regeneration [41], canonical differentiation paths detected in kidney organoids [42], and expected developmental programs in hepatic tissue. Hickey et al. [48] compared the gene expression (snRNA-seq) and accessibility (snATAC-seq) across pseudotime for different parts of the intestine, consistent with known functional differences.

**Identifying biologically-meaningful branching structures.** A third useful approach involves finding branching that and connecting it to biologically interpretable divergences, which is illustrated in Xin et al. [36], Kirita et al. [37], Oprescu et al. [41], Bao et al. [49], Chen et al. [44], Zhang et al. [45], Zhang et al. [46], Scripture-Adams et al. [50], and Zhou et al.[51] In several studies, branches correspond to clear distinctions between healthy or pathological states [36], [37], [41], [45], [46], [49], [50]. Kirita et al. [37] identified branches that separate trajectories between successful and failed cellular repairs after acute kidney injuries. Findings from Koenig et al. [52] found gene expression differences along pseudotime related to dilated cardiomyopathy across healthy and pathological states. When branching is supported by adequate intermediate cells and aligns with known biology, it becomes a powerful tool for identifying alternative fates, contrasting disease-associated states, and refining the understanding of how cells progress through complex transitions. Identifying such divergent structures has potential for clinical applications, as they provide a way of finding potential intervention targets.

**Multiple approaches.** Robust analyses used more than one approach paired with trajectory inference, such as using multiple trajectory inference methods or multiple data modalities to track a process. Mukherjee et al. [38] ensured that results were consistent between Monocle 2 and 3, Kannan et al. [40] used RNA velocity to establish directionality, Pichet Binette et al. [53] combined scRNA-seq with bulk proteomics to track progression of Alzheimer's disease. When multiple approaches converge on the same structure, the biological interpretation gains credibility. This use of complementary approaches shows how trajectory inference can be strengthened by cross validation instead of using the results from a single approach.

## Common pitfalls in literature

Published analyses across multiple domains reveal recurring challenges when interpreting results obtained from trajectory inference or pseudotime. We refrain from giving direct examples to avoid singling out individual papers, as the errors outlined here usually did not affect the overall quality of the work we would reference.

**Root choice sensitivity.** A few studies reported differences in pseudotime distributions between clusters, citing it as evidence of characteristic differences. Others point to the particular ordering of clusters ("before" or "after") without describing how the initial point was selected. Their conclusions are then a consequence of their selection of root cells and not of anything related to the biological process. Using the same reasoning, we could select the root cell at the end of the obtained trajectories and find a reverse order that would be equally-valid.

**Redundant trajectory inference.** Although this paper generally extols trajectory inference for single-cell data, its use is not always warranted. A number of publications produced trajectories that did not provide information that wasn't redundant with simpler analyses used in other parts of the papers. Differential expression between clusters, for instance, does not require trajectory inference. We suspect that producing those results is a consequence of reviewer pressure or as a way to add visually-appealing trajectory plots to a paper.

**Pseudotime as an independent measure of progression.** A recurring pattern is to interpret pseudotime as a direct readout of biological time or process stage without independent validation or calibration, even when external time-resolved measurements are available.

**Consistency as validation**. Validating a trajectory is important and having gene expressions along pseudotime that match known processes is a useful sanity check. Finding gene expression trends that are consistent with known biomarkers is treated as a complete validation, despite the problem of circularity, as described in Fang et al. (Suppl. 1) [24].

## Conclusion

In conclusion, trajectory inference has emerged as a powerful computational approach for unraveling the continuous and dynamic processes underlying cellular differentiation, development, and disease. By introducing different methods such as Monocle, Slingshot, PAGA, and Palantir, our review highlights how these tools not only enable the reconstruction of cellular trajectories from single-cell omics data but also pave the way for a multitude of downstream analyses—from gene regulatory network inference to cross-condition and cross-species comparisons. Yet, as the application examples and misuse cases illustrate, careful consideration of sampling adequacy, continuity assumptions, and the inherent limitations of pseudotime metrics is essential for avoiding overinterpretation. Moving forward, integrating robust external validation, multi-omic data types, and improved methodological frameworks will be crucial for refining these approaches. Ultimately, the continued evolution of trajectory inference will deepen our understanding of cellular behavior in both physiological and pathological contexts, offering promising insights into complex biological processes and disease mechanisms.

| Authors | Biological System or Pathology | Description | Omic layer | TI Method Data + Code |
|---|---|---|---|---|
| Xin et al. (2018) | Pancreas | Examining different states of pancreatic beta cells, specifically the impact of misfolding proinsulin on the transcriptome. | | Monocle 2.4<br>Data unavailable<br>Code unavailable |
| Kirita et al. (2020) | Acute kidney injury | Looked at the snRNA-seq trajectory of kidney cells after experiencing acute kidney injuries. | | Monocle 2<br>Data<br>Code unavailable |
| Mukherjee et al. (2020) | Alzheimer's Disease | Quantitative diagnosis of Alzheimer's. | | Monocle 2, Monocle 1, DPT<br>Data<br>Code |
| Lotto et al. (2020) | Liver | Cell composition in liver development. | | Palantir<br>Data<br>Code unavailable |
| Oprescu et al. (2020) | Fatty Infiltration | Investigated dynamics in cell populations in response to injury. | | Monocle<br>Data<br>Code unavailable |
| Bao et al. (2021) | Diabetes (Type 2) | Examined differences between the transcriptomic profiles of pancreatic Beta cells in healthy and diabetic patients. | | Monocle 2<br>Data unavailable<br>Code |
| Chen et al. (2021) | Gallbladder Cancer | Examined cell heterogeneity in the tumor microenvironment in gallbladder cancer. | | Monocle 2<br>Data†<br>Code unavailable |
| Zhang et al. (2021) | Gallbladder Cancer | Looked at enrichment of cell types between gallbladder tumor and its microenvironment. | | Monocle 2.14.0<br>Data unavailable<br>Code unavailable |
| Zhang et al. (2021) | Gallbladder Cancer | Examined the ErbB in the tumor environment of gallbladder cancer. | | Monocle 2<br>Data unavailable<br>Code unavailable |
| Smajić et al. (2022) | Parkinson's Disease | Profiled multiple cell types and contrasted omic profiles between healthy patients and those with Parkinson's. | | Monocle 3<br>Data*<br>Code unavailable |
| Scripture-Adams et al. (2022) | Muscular Dystrophy | Profiling transcriptomic differences in Duchenne muscular dystrophy mouse model. | | Slingshot<br>Data (1), (2)<br>Code unavailable |
| Yoshimura et al. (2023) | Kidney | Looked at gene regulatory networks in kidney organoids. | | Monocle 2<br>Data (1), (2), (3)<br>Code unavailable |
| Li et al. (2023) | Leukemia | Looked at cell population differences between categories of Leukemia to connect with prognosis. | 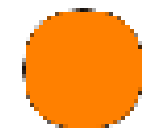 | Monocle 2<br>Data (1)†, (2)† |
| Pichet Binette et al. (2024) | Alzheimer's Disease | Looked at the presence of AD biomarkers in omics profiles. | 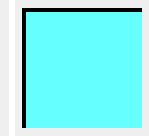 | Scorpius<br>Data unavailable<br>Code |
| Tosti et al. (2021) | Pancreas | Producing an atlas for human pancreas. | 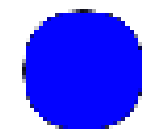 | PAGA<br>Data†<br>Code unavailable |
| Koenig et al. (2022) | Dilated cardiomyopathy | Studying transcriptional differences between healthy patients and those with DCM. | 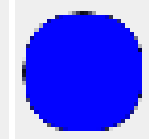 | Palantir<br>Data<br>Code |
| Zhou et al. (2022) | Pancreatic Cancer | Determine transition states of pancreatic cells in pancreatic ductal adenocarcinoma. | 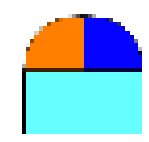 | Monocle 3<br>Data†<br>Code |
| Kannan et al. (2023) | Cardiomyocytes | Studying *in vivo* vs *in vitro* cardiomyocytes. | 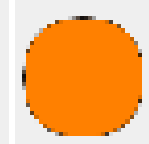 | Monocle 3<br>Data (1), (2)<br>Code |
| Hickey et al. (2023) | Intestine | Investigated cell differentiation in multiple parts of the intestine to show differences in their respective developments. | 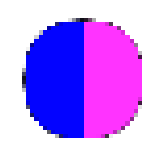 | ArchR<br>Data°<br>Code° |

**Table 1.** Summary of papers using applying trajectory inference.

†Data access requires application.

*Some data unavailable.

°Multiple sources directly available in paper.

**Table Legend:**

Circles indicate single-cell or -nucleus, squares indicate bulk, color indicates the omic layer.

| | | | |
|---|---|---|---|
| 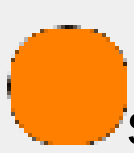 scRNA-seq | 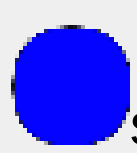 snRNA-seq | 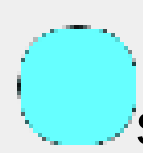 scP | 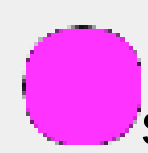 snATAC-seq |
| 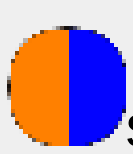 scRNA-seq + scP | 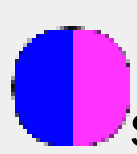 snRNA-seq + snATAC-seq | 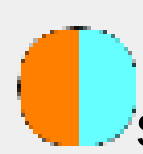 scRNA-seq + scP | |
| 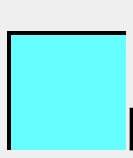 bulkP | 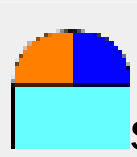 scRNA-seq + snRNA-seq + bulkP | 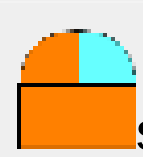 scRNA-seq + RNA-seq + scP | |

**Acknowledgements**
We thank Felice Information Design for graphic design help. This work was funded in part by the NIGMS (R35GM142502).